\documentclass[11pt]{article}
\usepackage{a4wide}
\usepackage{epsfig,epsf,graphics}
\usepackage{times}
\usepackage{authblk}
\usepackage{color}
\usepackage{amsfonts,amssymb,latexsym,amsmath} 
\usepackage{bm}
\usepackage{enumerate}
\usepackage{enumitem}
\usepackage{hhline}
\usepackage{longtable}
\usepackage{array} 
\usepackage[
colorlinks=true,
linkcolor=black,
breaklinks=true,
urlcolor=green,
citecolor=blue]{hyperref}
\usepackage{cite}
\usepackage{epstopdf}
\newcommand{\be}{\begin{equation}}
\newcommand{\eeq}{\end{equation}}
\newcommand{\ba}{\begin{eqnarray}}
\newcommand{\ea}{\end{eqnarray}}

\newcommand{\al}{&\!\!\!\!}

\newcommand{\ee}{e^+e^-}

\begin{document}
\title{Production cross section estimates  for strongly-interacting
Electroweak Symmetry Breaking Sector resonances\\ at particle colliders}

\author[1]{Antonio Dobado}
\author[2]{Feng-Kun Guo}
\author[1]{Felipe J. Llanes-Estrada}
\affil[1]{\it{\small Departamento de F\'{\i}sica Te\'orica, Universidad
Complutense de Madrid,}\authorcr
{\it\small Plaza de las Ciencias 1, 28040 Madrid, Spain}}
\affil[2]{\it{\small Helmholtz-Institut f\"ur Strahlen- und Kernphysik and Bethe
Center for Theoretical Physics,}\authorcr
{\it\small Universit\"at Bonn,  D-53115 Bonn,
Germany}}

\maketitle

\begin{abstract}

We are exploring a generic strongly-interacting Electroweak Symmetry Breaking
Sector (EWSBS) with the low-energy effectie field theory for the four
experimentally known particles ($W_L^\pm$, $Z_L$, $h$) and its
dispersion-relation based unitary extension. In this contribution we provide
simple estimates for the production cross section of pairs of the EWSBS bosons
and their resonances at proton-proton colliders as well as in a future $e^-e^+$
(or potentially a $\mu^-\mu^+$) collider with a typical few-TeV energy.
We examine the simplest production mechanisms, tree-level production through a
$W$ (dominant when quantum numbers allow) and the simple effective boson
approximation (in which the electroweak bosons are considered as collinear
partons of the colliding fermions). We exemplify with custodial isovector and
isotensor resonances at 2~TeV, the energy currently being discussed because of a
slight excess in the ATLAS 2-jet data.
We find it hard, though not unthinkable, to ascribe this excess to one of these
$W_LW_L$ rescattering resonances.
An isovector resonance could be produced at a rate smaller than, but close to
earlier CMS exclusion bounds, depending on the parameters of the effective
theory. The $ZZ$ excess is then problematic and requires additional physics
(such as an additional scalar resonance).
The isotensor one (that would describe all charge combinations) has a smaller
cross section.

\end{abstract}

\newpage

\section{Introduction}

If physics beyond the SM exists, the lack of any  manifestation in the
few-hundred GeV region and the lightness of the new Higgs-like boson naturally
suggest that this particle could be a quasi-Goldstone boson beyond the three
needed for Electroweak Chiral Symmetry Breaking.
 This would call for enlarging the Standard Model (SM) symmetry group, leading
 perhaps to composite Higgs models.

Independently of this, the current spectrum in the 100~GeV region consists of
the custodial-isospin triplet of $W^\pm$ and $Z$ bosons together with the new
Higgs boson $h$. A general formulation of the Electroweak Symmetry Breaking
Sector (EWSBS) in terms of effective field theory (in the non-linear realization
of $SU(2)_L\times SU(2)_R\to SU(2)_V$) can be encoded, neglecting boson masses, in
the seven-parameter next-to-leading order (NLO) Lagrangian density
\ba \label{bosonLagrangian} {\cal L}
& = & \frac{1}{2}\left[1 +2 a \frac{h}{v} +b\left(\frac{h}{v}\right)^2\right]
\partial_\mu \omega^i \partial^\mu
\omega^j\left(\delta_{ij}+\frac{\omega^i\omega^j}{v^2}\right) \nonumber
+\frac{1}{2}\partial_\mu h \partial^\mu h \nonumber  \\
 & + & \frac{4 a_4}{v^4}\partial_\mu \omega^i\partial_\nu \omega^i\partial^\mu
 \omega^j\partial^\nu \omega^j +
\frac{4 a_5}{v^4}\partial_\mu \omega^i\partial^\mu \omega^i\partial_\nu
\omega^j\partial^\nu \omega^j  +\frac{g}{v^4} (\partial_\mu h \partial^\mu h )^2
 \nonumber   \\
 & + & \frac{2 d}{v^4} \partial_\mu h\partial^\mu h\partial_\nu \omega^i
 \partial^\nu\omega^i
+\frac{2 e}{v^4} \partial_\mu h\partial^\nu h\partial^\mu \omega^i
\partial_\nu\omega^i
\ea
that we have described in detail
in Refs.~\cite{Delgado:2015kxa,Delgado:2013loa}.
(See also Refs.~\cite{Alonso:2014wta,Kilian:2014zja,Buchalla:2015wfa}
and references therein for additional background.)

The Equivalence Theorem (ET)~\cite{Cornwall:1974km} relates the amplitudes of
these $\omega$ Goldstone bosons (GBs) to those of the longitudinal components of
the electroweak gauge  bosons, $W_L$ and $Z_L$ in
 the SM and can also be extended to effective field theories~\cite{ETET} with
 larger particle/interaction content.

The effective Lagrangian of Eq.~(\ref{bosonLagrangian}) is useful in the 0.5--3
TeV region: for $E<0.5$ TeV the ET starts receiving large corrections, and for
$E>4\pi v\sim 3$ TeV the derivative expansion breaks down.  It includes the
newly found $h$ field coupled as an $SU(2)_V$ singlet in a custodially-invariant
way, but we are not concerned with it in this work, that concentrates on
$\omega\omega$ production with non-vanishing custodial isospin.
The reason for this focus is that the much commented ATLAS diboson
excess~\cite{Aad:2015owa}, barring misidentification, is seen in all $WW$, $WZ$
and $ZZ$ channels. A similar philosophy has been followed by the Barcelona
group~\cite{Espriu}.

If new resonances beyond the SM appear in the spectrum, the pure
(polynomial-like) momentum expansion fails before the $4\pi v$ scale as is
well-known from hadron physics, where there are elastic pion-pion resonances
below $4\pi f_\pi\simeq 1.2$ GeV.  The useful tools are then dispersion
relations, whose subtraction constants are fixed by the effective theory, so
that elastic (or coupled-channel, in the chiral limit) unitarity is exactly
enforced. In Appendix~\ref{sec:unitarization} we quickly review the resulting
Inverse Amplitude Method (IAM)~\cite{Truong:1988zp,IAM} that provides us with
$\omega\omega$ scattering amplitudes that are unitary, have the right analytic
properties for complex Mandelstam variable $s$, match perturbation theory based
on the Lagrangian in Eq.~(\ref{bosonLagrangian}) and are encoded in a very
simple algebraic formula, without the need for tedious numerical solutions of
involved integral equations.
Resonances can then be generated as poles of the unitarized $\omega\omega$
scattering amplitudes.

In this contribution we address the production cross section of exemplary
resonances generated by the IAM. We examine two production mechanisms, the
collision of two longitudinal $W_LW_L$ bosons collinear with the beam particles (effective boson approximation) in Section~\ref{sec:collinear}
yielding an isotensor $\omega\omega$ resonance,
and the production of an isovector one by an intermediate gauge boson in Section~\ref{sec:intermediateW}.

As will be shown in Section~\ref{sec:numerics}, the intermediate-$W$ boson
mechanism for the production of an isovector $\rho$-like resonance is larger
(since the isospin Clebsch--Gordan coefficients impede $\rho\not\!\!
\to\pi^0\pi^0$ the $ZZ$ data would be ascribed to misidentification or to a
concurrent scalar resonance as noted in~\cite{Arnan:2015csa}).
The computed cross section for the production of an isovector resonance (around
18~fb/TeV at 2~TeV) is just smaller than the related bounds provided by the CMS
collaboration in Ref.~\cite{Khachatryan:2014hpa} (about 20~fb/TeV there). We
conclude that an ATLAS excess with the same data base could only marginally be
generated by a resonance stemming purely from the EWSBS, though further more
detailed studies appear necessary.

Explanations invoking strong coupling of the new physics to quarks and gluons
have recently been proposed, but we do not address those.

\section{Cross section from collinear $\bm{W}$s}\label{sec:collinear}
\subsection{Lepton-lepton collisions\label{subsec:leptons}}

\begin{figure}[tb]
  \begin{center}
    \includegraphics[width=0.4\textwidth]{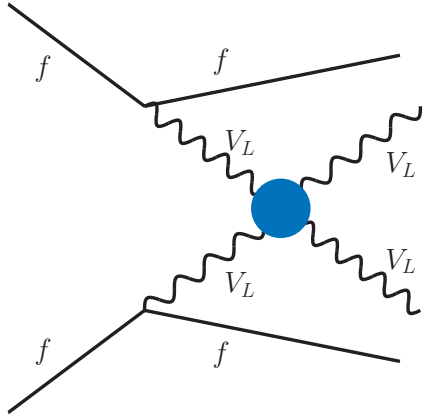}
    \caption{\label{fig:eeww} Production of a pair of large-$p_T$ longitudinal
    vector bosons by rescattering from two collinear $V_LV_L$-partons in a
    $e^+e^-$ collision (or generically, fermion-fermion collision such as
    quark-quark at the LHC). }
  \end{center}
\end{figure}

We start and settle notation with the effective $W$
approximation~\cite{Dawson:1984gx} in $e^-e^+$ collisions, that amounts to
treating the $V_L$ as a collinear parton of the lepton pair $e^-$ or $e^+$. Then
one can write down collinear factorization formula for the $e^+e^-\to
V_{L}V_{L}X$ process (with $X$ representing a pair of $e^+e^-$ or $\nu_e\bar
\nu_e$), as shown in Fig.~\ref{fig:eeww}, in terms of the parton-parton
($V_LV_L$ in this case) cross section.

The differential cross section for this production process as a function
of the $V_LV_L$ total center-of-mass energy $\sqrt{s}$ may be written
as~\cite{Dobado:1989ue}
\begin{equation}
  \frac{d\sigma}{ds} = \int_0^1 dx_+ \int_0^1 dx_- \,\hat
  \sigma(s)\, \delta(s -x_+x_-E_\text{tot}^2)\, \left[F_1(x_+)F_2(x_-) +
  F_2(x_-)F_1(x_+) \right]\ ,
  \label{eq:effW1}
\end{equation}
where inside the integral $\hat \sigma(s)$ is the cross section for the process
$V_{L1}V_{L2}\to V_{L3}V_{L4}$ with all the particles on-shell,
$E_\text{tot}$ is the center-of-mass energy of the initial, colliding pair of
$e^+e^-$, and $x_\pm$ are the energy fractions that the initial collinear
$V_L$'s take from their respective parent $e^\pm$ leptons. $F_{1,2}$ are
the lepton structure functions for $V_{L1,L2}$, and they were calculated
in Ref.~\cite{Dawson:1984gx} to be
\begin{equation} \label{pdfs}
  F_{W_L}(x) = g_{W} \frac{1-x}{x},\qquad
  	 F_{Z_L}(x) = g_{Z} \frac{1-x}{x},
\end{equation}
with
\begin{equation} \label{emissioncouplings}
   g_{W} = \frac{\alpha}{4\pi \sin\theta_W^2}, \qquad
   g_Z = \frac{\alpha[1+(1-4 \sin\theta_W^2)^2]}{16\pi \sin\theta_W^2 \cos\theta_W^2},
\end{equation}
$\alpha$ being the fine-structure constant and $\theta_W$ the Weinberg angle.
The $\delta$-function $\delta(s -x_+x_-E_\text{tot}^2)$ can be easily obtained
from $s=(p_1+p_2)^2$, $p_1=x_+ p_{e^+}$, $p_2=x_- p_{e^-}$ and
$E_\text{tot}^2=(p_{e^+}+p_{e^-})^2$ in the center-of-mass frame and neglecting
the lepton masses.

Noticing that $x_\pm$ are the lepton momentum fractions carried by the initial
vector bosons under the effective $W$ approximation, and they do not appear in
the vector-vector scattering cross section $\hat\sigma$ for fixed $s$, one can
factorize the cross section $\hat \sigma$ outside the integrations over $x_+$
and $x_-$.

We may then perform the integrations over the energy fractions analytically.
Once the $x_-$ integration has been carried out thanks to the $\delta$-function,
the lower limit of the $x_+$ integration becomes  $x_+\ge r$ with $r$ defined
as $r\equiv {s}/{E_\text{tot}^2}$\,, and we obtain a simple closed formula in
terms of the ratio $r$,
\begin{equation}
  \frac{d\sigma}{ds} = \frac{2}{s} g_1 g_2 \left[ 2 (r-1)-(r+1)\log r
  \right]  \hat \sigma(s),
  \label{eq:effW3}
\end{equation}
where the product $g_1 g_2$ is equal to $g_W^2 (g_Z^2)$ if the initial
vector mesons are $W_LW_L (Z_LZ_L)$ and $g_Wg_Z$ if they are $W_LZ_L$.
When $s\to E_\text{tot}^2$, $r\to 1$, we obtain a strong end-point suppression
(because it is unlikely that the vector boson takes a large momentum fraction of the lepton).
Moreover vector bosons at high energy are nearly transversely polarized because
of the strong Lorentz contraction.

The boson-boson cross section $\hat \sigma$ can be calculated using standard
formula for $2\to2$ cross sections given the scattering amplitude $A$.
It is convenient to obtain it in the center-of-mass frame of the vector boson pair,
\begin{equation}\label{cmcross}
  \frac{d\,\hat \sigma}{d\cos\theta} = \frac{S} {32\pi s}
  | A(s,\cos\theta)|^2\, ,
\end{equation}
where $\theta$ is the scattering angle. Then we convert it to a (longitudinal) reference-frame invariant cross section via the Mandelstam
variables as $\cos\theta=1+2\,t/s$ when masses for all particles are neglected (and for $\sqrt{s}\gg M_W$ we can
consider massless particles consistently with our use of the ET).
The symmetry factor $S$ in Eq.~(\ref{cmcross}) accounts for the identical particles in the final
state, and it takes the value of $1/2$ for the $Z_LZ_L$ case and $1$ for the
$W_L^+W_L^-$ case.

\subsection{Hadron colliders}\label{subsec:partons}

In the LHC context, the diagram in Fig.~\ref{fig:eeww} represents the production
in elementary quark-quark collisions, so the parton distribution functions (pdfs) of Eq.~(\ref{pdfs}) (also related to the luminosity functions for $V_L$ splitting from quarks) describe the probability of finding a longitudinal boson splitting collinearly from a quark/antiquark. The only difference is in the auxiliary coupling $g_Z$ of Eq.~\eqref{emissioncouplings}, because of the
different isospin and hypercharges for the up and down-type quarks. This changes the respective coefficient of $\sin\theta_W^2$ as follows,
\ba
\label{couplingquark}
g_{Z}^u = \frac{\alpha[1+(1- \frac{8}{3} \sin\theta_W^2)^2]}{16\pi \sin\theta_W^2 \cos\theta_W^2},
\qquad
g_Z^d = \frac{\alpha[1+(1-\frac{4}{3} \sin\theta_W^2)^2]}{16\pi \sin\theta_W^2 \cos\theta_W^2}\ .
\ea

Now we can construct the wanted pdf for the vector boson in the proton by
convolving the one in the quark with the pdf of the quark on the proton itself.
This is~\cite{Dawson:1984gx} \be
\label{Wintheproton}
F_{W_L}^p (x) \equiv \int_x^1 \frac{dy}{y} \sum_i f_i(y)\times F_{W_L}^{q_i} \!
\! \left(\frac{x}{y}\right) \ .
\eeq
The $y$ variable swipes the momentum fraction of the emitting quark in the
proton, distributed according to $f_i(y)$, and that quark propagator is $1/y$.
The flavor index $i$ traverses ten quark/antiquark flavors ($u$, $d$, $s$, $c$,
$b$ and their antiquarks). The only flavor dependence other than $f_i$ is in the emission coupling for the $Z$ boson in Eq.~(\ref{couplingquark}). Finally, $x$ is the momentum fraction of the vector boson inside the proton, and takes values
in the interval $x\in(M_W/E_{\rm proton},1)$.
The $Z$-boson is treated in the same way, replacing $M_W$ with $M_Z$ and writing
down an equation analogous to Eq.~(\ref{Wintheproton}).

For the pdf of the quark $f_i(x)$ we resort to the well-known and widely used
CTEQ set; we take their last issue, the CJ12 distributions with maximum
nuclear and $Q^2$ corrections~\cite{Owens:2012bv}. We have checked that using
other corrections has a very little impact on the cross section estimates.

\section{Cross section from intermediate gauge boson production}
\label{sec:intermediateW}
In this section we provide a quick estimate for the cross section $\sigma(pp\to
W+X \to wz+X)$ where the GB pair $ww$ is (through ET)
interchangeable for $W_LW_L$, and we take into account the rescattering of the
final state bosons (which makes the calculation not totally trivial).

The reason for choosing the $w z$ channel for the illustration is because the
ATLAS excess is possibly seen (if not a misidentification) in the charged $WZ$
dijet spectrum.

The leading tree-level amplitude for the process must come then from the
annihilation of the lightest $q\bar{q}$ pair with total unit charge, namely
$u \overline d \rightarrow W^+ \rightarrow w^+ z$, and is given by
\be
T(s,\theta,\phi)= \frac{g^2}{2 \sqrt 2}\sin \theta e^{- i \phi},
\label{eq:a1w}
\eeq
This amplitude is purely $J=1$ corresponding to a negative helicity $u$ and a
positive helicity $\overline d$.

\begin{figure}
\centerline{\includegraphics[width=6cm]{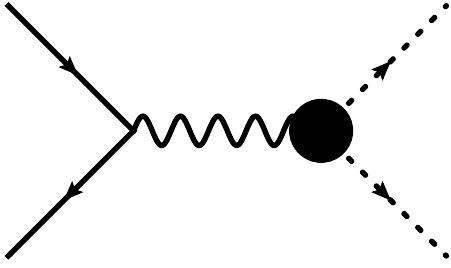}}
\caption{Tree-level GB production via the annihilation of a $u\bar{d}$ quark
into a gauge $W^+$ boson. Strong rescattering in the final state appears through the form factor $F_V(s)$ represented by the thick blob.}
\label{fig:tree}
\end{figure}

The rescattering of the final $w^+ z$ would-be GBs can be
taken into account easily by introducing the vector form factor $F_V(s)$ of
Eq.~\eqref{ffactor} below in agreement with Watson's final state theorem. This
form factor, the thick blob in the Feynman diagram of Fig.~\ref{fig:tree},
compactly encodes all the strong GB dynamics in this channel, eventually
including a vector resonance. As it was shown in
Ref.~\cite{Truong:1988zp,Dobado:1999xb} it is possible to use the IAM method
(see Appendix A) to obtain this form factor in terms of the $I=J=1$ partial wave as obtained from the one-loop effective theory
to find:
\begin{equation}
F_V(s) = F_{11}(s)
      = \left[1-\frac{A_{11}^{(1)}(s)}{A_{11}^{(0)}(s)}\right]^{-1} .
\end{equation}
where $A_{11}^{(0)}(s)$ and $A_{11}^{(1)}(s)$ are the tree-level and one-loop
contributions to the partial wave.

The unpolarized center-of-mass cross section is then
\begin{eqnarray} \label{udbarWW}
\frac{d \hat{\sigma}(u \overline d \rightarrow w^+ z)}{d \Omega_\text{CM}} =
\frac{1}{64 \pi^2 s} \left( \frac{1}{4} \right) \left( \frac{g^4}{8} \right)
\mid F_V(s)\mid^  2 \sin ^2 \theta
\ .
\end{eqnarray}
Note that an identical formula can be used for the reaction $d \overline u
\rightarrow W^- \rightarrow w^- z$, and  that we are neglecting masses and
Cabbibo--Kobayashi--Maskawa mixing.
In principle these subprocesses are formally suppressed with respect to the pure
GB elastic scattering in this channel (longitudinal gauge boson
fusion) whose amplitude is given by
\be
T(ww \rightarrow ww) = 96 \pi\, \cos \theta\, A_{11}(s)\ ,
\eeq
where we have truncated at the $J=1$ partial wave, and $A_{11}(s)$ is the
$J=I=1$ partial wave for $ww$ elastic scattering (see Appendix A). It is of
order $O(1)$ instead of $O(\alpha)$ found in Eq.~\eqref{eq:a1w}.

However, this is again only the parton-level process. In the LHC environment, we
need to take the parton distribution functions into account, and here a pair of
$u\bar{d}$ fermions are more readily available than $W_LW_L$.
It turns out that this process is dominant as will be shown numerically below in
Section~\ref{sec:numerics}.

Convolving Eq.~(\ref{udbarWW}) with the pdfs $f(x)$ as
described earlier in Section~\ref{subsec:partons}, we obtain the proton-proton
inclusive cross section to produce a pair of GBs as
\be
\label{ppintermediateW}
\frac{d\sigma}{ds}(pp\to w^+z + X) =
\int_0^1 dx_{u} \int_0^1 dx_{\bar d}\, \delta\left(s-x_u x_{\bar
d}E_\text{tot}^2\right)  \hat\sigma(u\bar d \to w^+z) f(x_u) f(x_{\bar d})\, ,
\eeq

To conclude this section, let us note that in the limit of vanishing
hypercharge $g'=0$, custodial symmetry predicts a few relations
\begin{eqnarray}
\frac{d \hat{\sigma}(u \overline d \rightarrow w^+ z )}{d \Omega_\text{CM}}&=&
\frac{d \hat{\sigma}(u \overline u \rightarrow w^+w^- )}{d \Omega_\text{CM}}
\nonumber \\ &=&
\frac{d \hat{\sigma}(d \overline d \rightarrow w^+w^- )}{d \Omega_\text{CM}}
\nonumber \\
 &=& \frac{d \hat{\sigma}(e^+ e^- \rightarrow w^+w^- )}{d \Omega_\text{CM}} \, ,
\end{eqnarray}
so that our numerical computation for the reaction in Eq.~(\ref{udbarWW})
can be immediately used to estimate several others.

\section{Numerical results and discussion} \label{sec:numerics}

\subsection{Parameters}

The Weinberg angle in Eq.~(\ref{emissioncouplings}) corresponds to the
tree-level radiation of a gauge boson, so it can be
taken~\cite{Agashe:2014kda,Marciano:2000yj} as $\sin^2 \theta_W =0.231$ (at the
next order one should use the $\overline{\rm MS}$ value at the $M_Z$ pole, but
this higher precision is irrelevant for us). Likewise, we take
$\alpha(M_Z)\simeq 1/129$. With this, the auxiliary couplings in
Eq.~(\ref{emissioncouplings}) are determined to be about $g_W\simeq
2.67\times10^{-3}$ and $g_Z=8.73\times10^{-4}$.

\begin{figure}[tbh]
\centerline{\includegraphics*[width=0.6\textwidth]{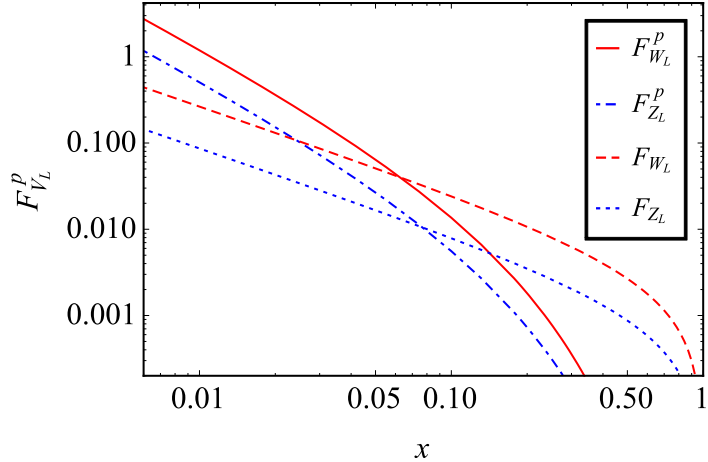}}
\caption{$W_L$ and $Z_L$ parton distribution functions in the proton (solid
and dot-dashed curves), employing the simple low-$x$
formula Eq.~(\ref{Wintheproton}) at a 6.5 TeV proton energy, and the electron
(dashed and dotted curves), using Eq.~\eqref{pdfs}.
\label{fig:pdfproton}}
\end{figure}

Once the generic parameters have
been fixed, we can obtain the pertinent gauge boson--parton distribution
functions in the effective boson approximation.
The ones for the $e^+e^-$ collisions, $F_{W_L}$ and $F_{Z_L}$ from
Eq.~(\ref{pdfs}), are shown as the dashed and dotted curves in
Fig.~\ref{fig:pdfproton},
and those appropriate for a 6.5~TeV proton beam (the LHC run II operates at
13~TeV in center-of-mass energy) are shown as solid and
dot-dashed curves in the same figure.
One can clearly see that, at the same energy, it is more likely to split a
vector boson from the proton at low $x$, and less likely at moderately high $x$
(since the quark pdfs in the proton typically fall off as $(1-x)^3$).

Moving now to the parameters of the effective Lagrangian density in
Eq.~\eqref{bosonLagrangian}, the concurrent constraints on the value of $a$ from
CMS and ATLAS~\cite{ATLAS:2014yka} indicate, at $2\sigma$, that $a\in
(0.88,1.3)$, that is, around the Standard Model value 1, so that the leading
order (LO) interaction strengths in the $IJ=00$, $11$ and $20$ channels, being
proportional to $\pm(1-a^2)$, are small and do not produce
elastic-$\omega\omega$ dynamically-generated states easily (inelastic
$\omega\omega-hh$ are much more unconstrained as observed
in Ref.~\cite{Delgado:2015kxa}).

\begin{figure}[tbh]
\centerline{
\includegraphics*[width=0.495\textwidth]{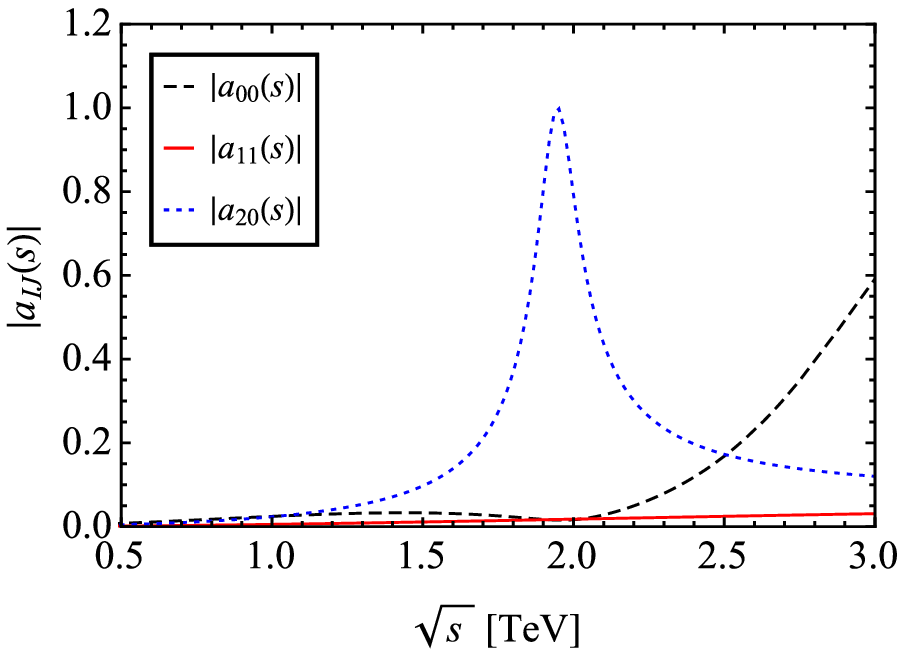}\hfill
\includegraphics*[width=0.495\textwidth]{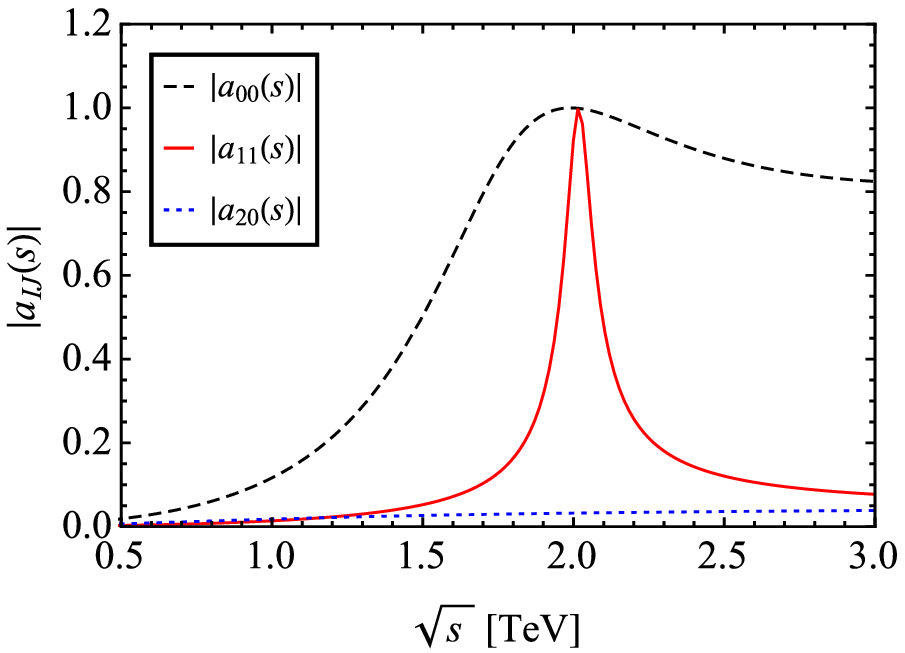}}
\caption{
\label{fig:amp} Moduli of the $\omega\omega\to \omega\omega$ amplitudes in
different spin-isospin channels unitarized using the IAM. Left: a narrow
scalar-isotensor resonance around 2~TeV is generated with the first parameter set.
Right:
a narrow vector-isovector resonance and a broad scalar-isoscalar resonance
around 2~TeV are generated with the second parameter set. }
\end{figure}

We resort to the NLO couplings to induce resonances in the unitarization
process, taking as a first set $a=1.05$, $b=1$, $a_4=1.25\times10^{-4}$ at a
scale $\mu=3$ TeV, and as a second set $a=0.9$, $b=a^2$, $a_4=7\times 10^{-4}$
(also at $\mu=3$ TeV), with all other couplings set to zero. The first set
produces an exemplary narrow isotensor resonance at around 2~TeV~\footnote{Note
that for this set $a>1$ and the QCD-like repulsive nature of the isotensor
channel is reversed, so an isotensor pole is possible, while an isovector one
becomes more difficult and violates causality in much of parameter space, see
Fig.~22 of Ref.~\cite{Delgado:2015kxa}.}
and the second set produces a narrow vector-isovector resonance (akin to a $W'$
or a Higgs-composite model $\rho$~\cite{Barducci:2015oza}) and a broad
scalar-isoscalar one, both of which are around 2~TeV. Theses exemplary
resonances can be clearly seen in the moduli of the amplitudes shown in
Fig.~\ref{fig:amp} (for explicit expressions of these amplitudes, we refer to
Ref.~\cite{Delgado:2015kxa}).

From the parameter space of the effective field theory reported
in Ref.~\cite{Delgado:2015kxa} we have chosen these two sets because the
resonances generated have a mass close to 2 TeV, so they would be clear
candidates to explain the putative ATLAS resonances.

\subsection{Estimate of the cross sections}

\begin{figure}[tbh]
  \begin{center}
    \includegraphics[width=0.495\textwidth]{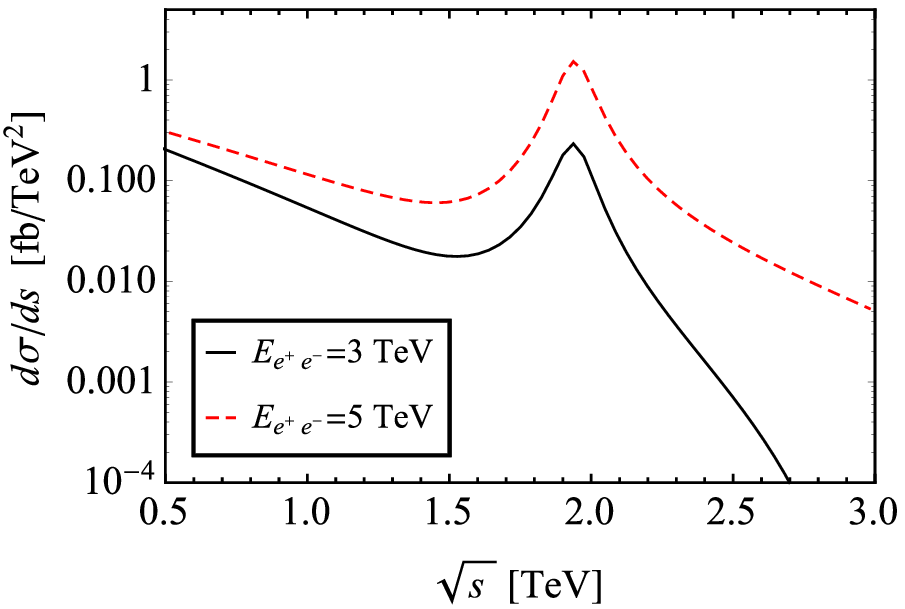}\hfill
    \includegraphics[width=0.495\textwidth]{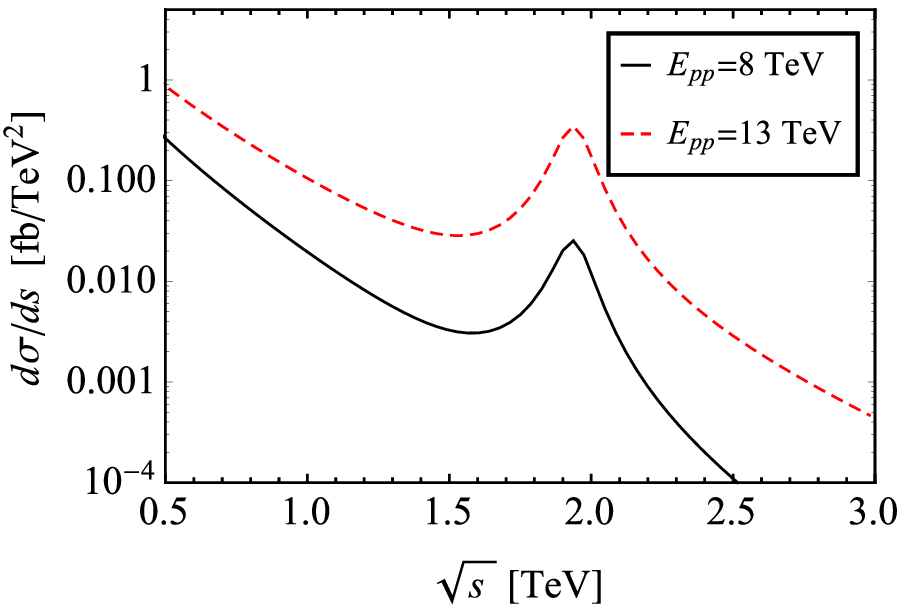}
    \caption{\label{fig:xeewwisotensor} Differential cross section for the
    production of a pair of $\omega\omega$ in $e^+e^-$ (left) and $pp$ (right)
    collisions with the effective boson approximation. Here we use $a=1.05$, $b=1$, $a_4=1.25\times10^{-4}$, and all the other couplings are set to zero (with $\mu=3$ TeV). This produces an IAM scalar-isotensor interaction.}
  \end{center}
\end{figure}

First, let us see what the effective boson approximation of
Section~\ref{sec:collinear} produces for the case of an isotensor resonance.
In Fig.~\ref{fig:xeewwisotensor}, we show the differential cross section for the
production of a pair of $W_L^+W_L^-$ in both electron-positron and proton-proton
collisions. We have summed up the individual cross sections with $W_L^+W_L^-$
and $Z_LZ_L$ in the initial state. We use the parameter set that generates an
isotensor resonance (able to simultaneously explain an excess in all $WW$, $WZ$
and $ZZ$ channels) which is visible in the curves.
One sees that the peak differential cross section at the LHC run-I with a 8~TeV
total energy is well below 0.1~fb/TeV$^2$. It is increased by one order of
magnitude at the 13~TeV LHC run-II operational energy and at a 3~TeV
electron-positron collider, and reaches 1~fb/TeV$^2$ at a 5~TeV lepton
collider.

\begin{figure}
\centerline{
\includegraphics[width=0.495\textwidth]{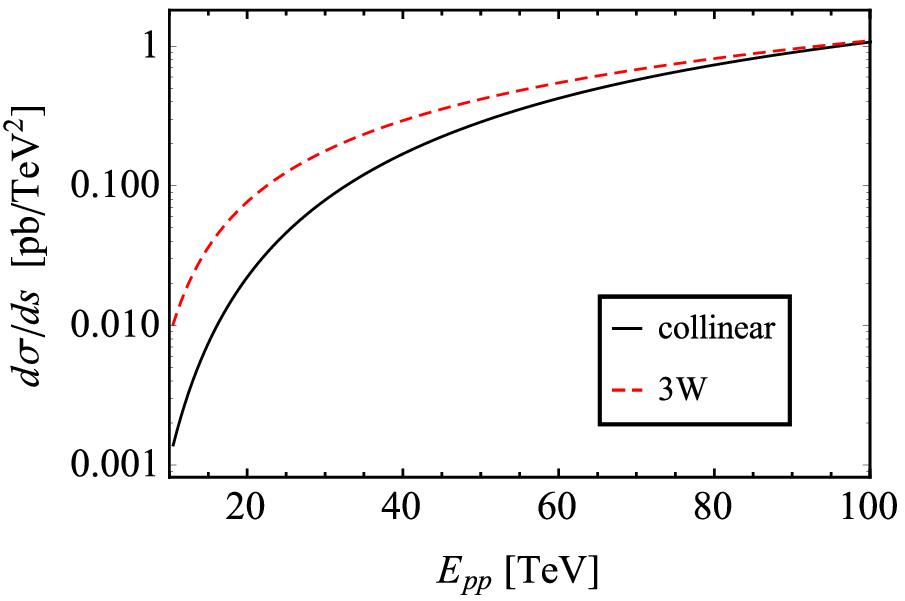}\hfill
\includegraphics[width=0.495\textwidth]{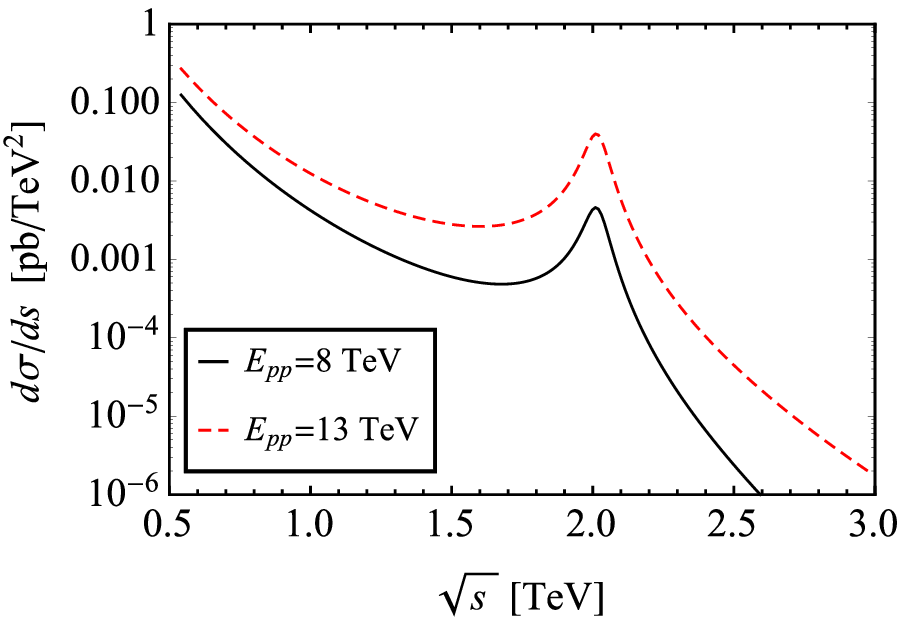}}
\caption{\label{fig:nonres}
Left: dependence of the differential cross section for the inclusive production
of a pair of isovector $W_L^+W_L^-$ at the peak of the isovector resonance on
the total proton-proton energy.
Here, the mechanism of the effective boson approximation is denoted by ``collinear'', and that through an
intermediate $W$ boson is denoted by ``3W''.
Right:
cross section for $pp\to W_L^+ Z_L +X$ through an intermediate $W^+$ in the presence of strong final-state interactions that induce a resonance in the channel with $J=1$ and $I=1$.}
\end{figure}

Next, we turn to the case of an isovector resonance. In this case, the mechanism
shown in Fig.~\ref{fig:eeww} is much less important than the mechanism,
described in Section~\ref{sec:intermediateW}, of an intermediate $W$ boson. This
can be clearly seen in the left panel of Fig.~\ref{fig:nonres} which
was calculated using the second parameter set.
%
The right panel of Fig.~\ref{fig:nonres} shows the inclusive cross section in
the proton-proton collisions through an intermediate $W$ boson in the presence
of a vector-isovector resonance generated using the second parameter set. If we
switch off the resonance, i.e., with $F_V(s)=1$, the cross section will drop
exponentially without any enhancement at around 2~TeV.

\begin{figure}[tbh]
\centerline{\includegraphics[width=0.6\textwidth]{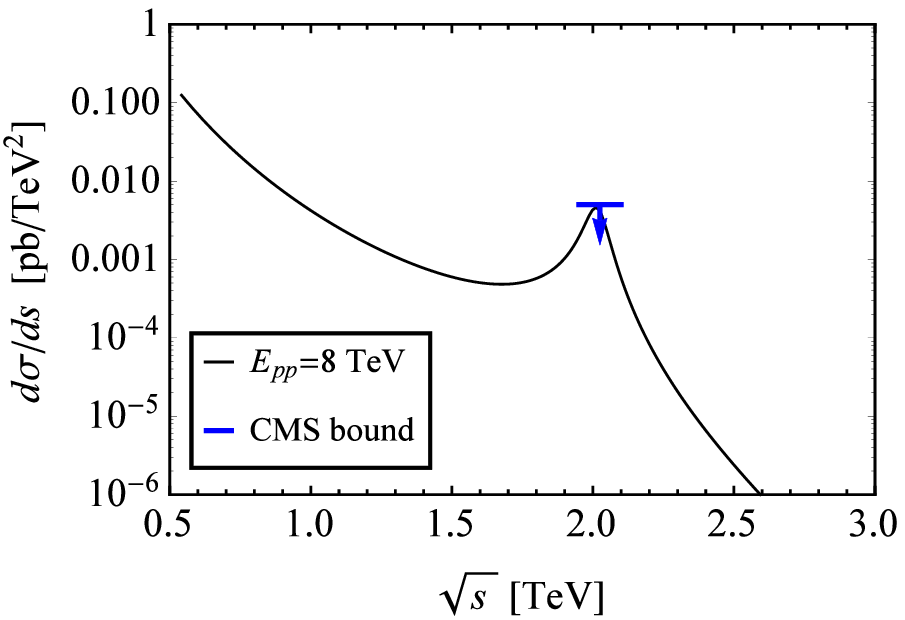}}
\caption{\label{fig:withCMS}  Tree-level $W$ production of $\omega\omega$ from
Eq.~(\ref{ppintermediateW}) in the presence of resonant final-state
interactions. Also shown is the CMS upper bound on the cross section for
production of a $W'$-like boson at 2 TeV (Fig. 6, left plot,
of~\cite{Khachatryan:2014hpa}, that we divided by $2E$ to convert $d\sigma/dE$
to $d\sigma/ds$).}
\end{figure}
The peak cross section for $E_{pp}=8$~TeV is about
$d\sigma/ds\simeq4.6$~fb/TeV$^2$ or $d\sigma/dE\simeq18$~fb/TeV.
As shown in Fig.~\ref{fig:withCMS}, it is very close to the CMS upper bound on
the production cross section, about 20~fb/TeV, under the assumption of the
resonance being an isovector $W'$ boson (alternative assumptions in Ref.~\cite{Khachatryan:2014hpa} are not too different).

\subsection{Summary and conclusions}

It appears that the expected production rate of resonances stemming purely from
the EWSBS is near and below the CMS reach with the statistics accumulated in run
I at 8 TeV (see Fig.~\ref{fig:withCMS}). We do find parameter sensitivity. For
example, if the values of $a=0.9$, $a_4=7\times 10^{-4}$ are modified to
$a=0.88$, $a_4=8\times 10^{-4}$, the cross section at the (approximately) 2 TeV
peak drops by a factor 2, and falls way below CMS's exclusion reach.

It then remains hard to believe, though open, that the ATLAS excess at 2~TeV in
the diboson channel can be attributed to purely EWSBS-resonances alone.
Our argumentation is rather model-independent as we rely on unitarized effective
field theory  without commitment to specific underlying BSM
mechanisms.~\footnote{Using different unitarization methods can result in some
model dependence, however, the glossary features for the dynamically generated
electroweak resonances remain the same as discussed in
Ref.~\cite{Delgado:2015kxa}.}

On the other hand, we have not examined fermion couplings, and new physics that
couples intensely to the QCD partons in the initial state remains an option as
the cross section would be increased respect to the $\alpha$-suppressed rate to
produce an intermediate $W$ boson. Low-energy flavor tests however challenge
such an interpretation, as remarked by other authors.
We are currently executing an extended investigation of the generic EWSBS sector
coupled to fermions in a symmetry-respecting effective Lagrangian in the
framework of another collaboration. Another alternative interpretation of the
data has been recently proposed~\cite{Aguilar-Saavedra:2015rna} in which an
additional boson has escaped detection (in spite of the already large cross
section).

Run II at 13 TeV  will improve the situation regarding the exclusion of purely
electroweak-symmetry breaking sector resonances because the cross section
(largely thanks to much increased parton luminosity) will increase
substantially, as seen in Fig.~\ref{fig:nonres}. Right now, there is just not
enough sensitivity.
Another interesting way of increasing the cross section, as we showed in
Fig.~\ref{fig:xeewwisotensor}, is to proceed to a lepton collider where the
initial state pointlike fermions are much more energetic, or to construct a
higher-energy hadron collider (the longitudinal $W_LW_L$ production mechanism
becomes competitive around 100 TeV) such as the second phase of the proposed
Circular Electron-Positron Collider--Super Proton-Proton Collider
(CEPC-SPPC)~\cite{CEPC}.

For the time being, we conclude that longitudinal $W_LW_L$ collinear radiation
is not a competitive production mechanism at present energies, becoming
important for an $O(80-100)$TeV pp collider, and that the ATLAS excess, if not a
statistical fluctuation as the collaboration keeps as working hypothesis, does
not easily fit as a resonance purely coupled to the electroweak gauge bosons,
rather independently of model considerations.

\medskip

\section*{Acknowledgments}

We warmly thank intense discussions and information exchange with J. J. Sanz
Cillero and  D. Espriu. We are grateful to U.-G.~Mei\ss{}ner for a careful
reading of the manuscript. This work is partially supported by the Spanish
Excellence Network on Hadronic Physics FIS2014-57026-REDT, by grants UCM:910309, MINECO:FPA2011-27853-C02-01,
MINECO:FPA2014-53375-C2-1-P, by DFG and NSFC through funds provided to the
Sino-German CRC 110 ``Symmetries and the Emergence of Structure in QCD" (NSFC Grant No. 11261130311) and by NSFC (Grant No. 11165005).

\begin{appendix}

\section{Strongly interacting amplitudes and form factors}

\subsection{Isospin relations} \label{subsec:isospin}

The isospin and partial wave expansions for the $\omega\omega$ scattering
amplitudes in the isospin basis $A_I(s,t,u)$ can be found
in~\cite{Delgado:2015kxa}; here we show a few equations of interest.

For the process $\ee\to \ee W_L^+W_L^- $, the initial vector bosons are
$Z_LZ_L$. Thus, the relevant rescattering process is $Z_LZ_L\to W_L^+W_L^-$, whose amplitude is given by
\begin{equation}
  A_{zz\to w^+w^-}(s,t,u) = \frac13 \left[ A_0(s,t,u) - A_2(s,t,u)
  \right],
\end{equation}
which can be easily obtained from the isospin relations noticing $A_{w^+w^-\to
zz}(s,t,u) = A(s,t,u)$.
The initial vector bosons are $W_L^+W_L^-$ for the process $\ee\to
\nu_e\bar\nu_e W_L^+W_L^- $, and we have
\begin{equation}
  A_{w^+w^-\to w^+w^-}(s,t,u) = \frac16 \left[ 2A_0(s,t,u) + 3A_1(s,t,u) +
  A_2(s,t,u) \right].
\end{equation}
While for the processes $\ee\to \ee Z_LZ_L $ and  $\ee\to
\nu_e\bar\nu_e Z_LZ_L $, we have the amplitudes
\begin{equation}
  A_{zz\to zz}(s,t,u) = \frac13 \left[ A_0(s,t,u) + 2A_2(s,t,u)
  \right]
\end{equation}
and
\begin{equation}
  A_{w^+w^-\to zz}(s,t,u) = \frac13 \left[ A_0(s,t,u) - A_2(s,t,u)
  \right],
\end{equation}
respectively. If the vector boson pair is charged, i.e. $W_L^\pm Z_L$, there is
no contribution from the isospin scalar channel. For such a process as $\ee \to
\bar\nu_e e^- W_L^+ Z_L$, the relevant scattering amplitude is
\begin{equation}
  A_{w^+z\to w^+z}(s,t,u) = \frac12 \left[ A_1(s,t,u) + A_2(s,t,u)\right].
\end{equation}

These scattering amplitudes are related to the partial wave ones by
\begin{equation}
  A_I(s,t,u) = 16 N\pi \sum_{J=0}^\infty (2J+1) P_J(\cos\theta) a_{IJ}(s),
\end{equation}
where $N=2$ if all the particles in the initial and final states are identical,
and $N=1$ otherwise. The unitarized expressions for the partial wave amplitudes
are shown next in Appendix~\ref{sec:unitarization}.

If we truncate the summation over $J$ at $J=2$, the invariant amplitudes can be
reconstructed easily from the partial waves by
\begin{eqnarray}
  A_0(s,t,u) \al=\al 16 N\pi\, \left[a_{00}(s) + \frac12
  \left(3\cos^2\theta-1\right) a_{02}(s) \right], \nonumber\\
  A_1(s,t,u) \al=\al 48 N\pi\, a_{11}(s) \cos\theta, \nonumber\\
  A_2(s,t,u) \al=\al 16 N\pi\, a_{20}(s) \ .
\end{eqnarray}

\subsection{Unitarization procedure: IAM}
\label{sec:unitarization}

In this section, we will briefly describe our unitarization procedure, the Inverse Amplitude Method  (IAM) \cite{IAM, Delgado:2015kxa}.

The effective-theory, partial-wave projected amplitudes satisfy on their
right-hand cut (RC) unitarity only perturbatively, reading ${\rm Im}\,  A^{(1)}
= (A^{(0)})^2$ with $(0)$ and $(1)$ denoting LO and NLO only, respectively. This
follows easily from their generic structure
\be
A^{(0)}(s)=K s, \qquad A^{(1)}(s) = \left[B(\mu)+D \log\frac{s}{\mu^2}+E
\log\frac{-s}{\mu^2}\right] s^2 \, ,
\eeq
and the field theory computation of the constants $B$, $D$ and $E$.

A complex-$s$ analysis of the elastic partial-wave scattering amplitude $A(s)$
yields an exact, but not too useful, dispersion relation for $A(s)$, and that  for $A^{(1)}(s)$ is not necessary because it is known everywhere from perturbation theory.
A useful technique is to apply a dispersive analysis to the following auxiliary
 function, \be \label{gdef}
  w(s)\equiv \frac{(A^{(0)}(s)) ^2}{A(s)}\ .
\eeq
   This $w(s)$ has the same analytic structure as $A(s)$
but for poles (at the zeroes of $A(s)$) that have been treated in the past~\cite{GomezNicola:2007qj} and concluded to be irrelevant for the physical region of $s$.
 Moreover, $w(0)=0$, $w(s)= K s  + O (s^2)$, and on the RC one has  ${\rm Im}\,
 w(s)=-(A^{(0)}(s)) ^2$. The twice-subtracted dispersion relation for this
 function, sufficient for one-channel problems, reads
\be
w(s)=K s+\frac{s^2}{\pi}\int_0^{\Lambda^2}  \frac{ds' {\rm Im} w(s')}
{s'^2(s'-s-i\epsilon)} +\frac{s^2}{\pi}\int_{-\Lambda^2}^{0}   \frac{ds' {\rm
Im} w(s')}{s'^2(s'-s-i\epsilon)}  + \frac{s^2}{2 \pi i}\int_{C_\Lambda}
\frac{ds' w (s')}{s'^2(s'-s)} \, ,
\eeq
where $\Lambda$ is a ultraviolet cutoff.
With the definition of $w(s)$ given in Eq.~\eqref{gdef}, one can compute
the elastic-RC integral \emph{exactly} since ${\rm Im}\, w(s)=-K^2s^2= E \pi
s^2$ there. This is dominant because it is the nearest complex-plane singularity
to the physical boundary which is the upper edge of the RC in the first Riemann
sheet.

Because the left-hand cut (LC) integral cannot be obtained exactly, it is
customarily computed in perturbation theory.
As discussed in Ref.~\cite{Delgado:2015kxa}, it is a very
reasonable approximation to take
\be
{\rm Im} w(s) \simeq - {\rm Im} A^{(1)}(s) \, ,
\eeq
which leads to
 \be
w(s) \simeq K s-D s^2\log\frac{s}{\Lambda^2}-E s^2\log\frac{-s}{\Lambda^2} +
\frac{s^2}{2 \pi i}\int_{C_\Lambda}  \frac{ds'  w(s')}{s'^2(s'-s)}  \ .
\eeq
This approximate integral equation is solved by
$w(s)=  A^{(0)}(s)- A^{(1)}(s)$.
In the above derivation, the only used approximations are the absence of poles
in $w(s)$ and the perturbative treatment of the LC integral.
Therefore, from the definition of the $w(s)$ in Eq.~\eqref{gdef} we get the
partial-wave amplitude in IAM as
\be \label{IAM1channel}
 A(s) \simeq  A^\text{IAM}(s)  =   \frac{(A^{(0)}(s))^2}{A^{(0)}(s)-A^{(1)}(s)}
 \ .
\eeq
This IAM amplitude has the proper analytic structure and makes poles on the
second Riemann sheet possible which correspond to dynamically generated
resonances. Elastic unitarity is satisfied by construction, and the amplitude is
also scale independent.
Furthermore, expanding at low energies, the IAM amplitude coincides with the one
in chiral perturbation theory up to NLO,
\be
A^\text{IAM}(s) =A^{(0)}(s) + A^{(1)}(s) + O(s^3) \, .
\eeq

\begin{figure}
\centerline{
\includegraphics[width=0.6\textwidth]{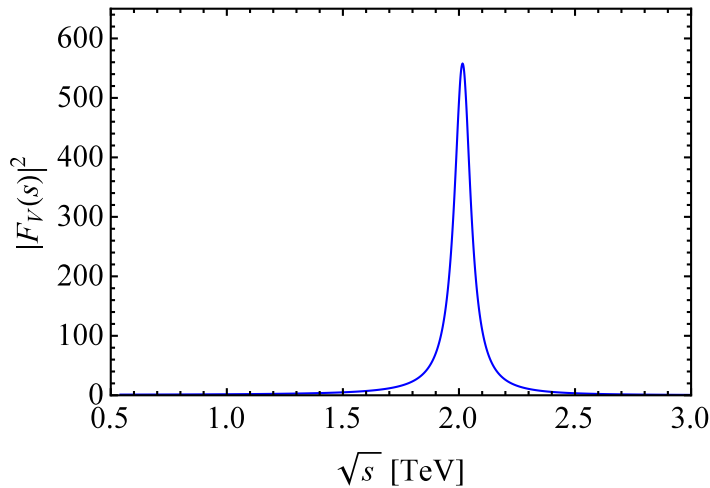}}
\caption{\label{fig:FF} Vector-isovector form factor with a narrow resonance at
about 2 TeV.}
\end{figure}

Watson's final state theorem~\cite{Watson:1952ji}
guarantees that the phase of the form factor for $W\to \omega\omega$
represented as the black blob in Fig.~\ref{fig:tree} is the same as that of
the elastic $\omega\omega$ scattering amplitude, and any resonance pole of the
scattering amplitude also appears in the form factor at the same position in
the complex $s$-plane.
Together with the normalization of the vector form factor $F_V(0)=1$ we find
that the form factor consistent with the IAM is given by
\begin{eqnarray} \label{ffactor}
F_V(s) = F_{11}(s)
       = \left[1-\frac{A_{11}^{(1)}(s)}{A_{11}^{(0)}(s)}\right]^{-1} .
\end{eqnarray}
This construction agrees with the perturbative expansion, has the correct unitarity cut, and shares phase with the corresponding scattering amplitude in the same $11$ channel.
Figure~\ref{fig:FF} shows the vector-isovector form factor necessary for
Eq.~(\ref{ppintermediateW}) with the parameter set
$a=0.9$,  $b=a^2$, $a_4=7\times 10^{-4}$.

\section{Kinematics of the effective boson approximation}
\label{app:kinetic}
\renewcommand{\theequation}{\thesection.\arabic{equation}}
\setcounter{equation}{0}

In this appendix we collect some useful relations among the kinematic variables of section~\ref{sec:collinear}.
Specifically, we relate the transverse momenta of the
vector mesons in the final state to the Mandelstam variables and the
center-of-mass scattering angle $\theta$ which appear in the scattering
amplitudes.

Let us start from the Mandelstam variable $t\equiv (p_1-p_3)^2$ for
the two-body scattering process $V_1(p_1)V_2(p_2)\to V_3(p_3)V_4(p_4)$.
For the case $m_1=m_3$ and $m_2=m_4$,
\begin{eqnarray}
  t = -2 {\bf p }_\text{cm}^2 (1 - \cos\theta),
  \label{eq:t1}
\end{eqnarray}
where ${\bf p}_\text{cm}$ is the modulus of the momentum in the
center-of-mass  frame of the initial (or final) state.

On the other hand, we can decompose ${\bf p}_3^{\,*}$, which is the momentum for
particle $V_3$ in the center-of-mass frame of the $V_3V_4$ system, into ${\bf
 p}_3^{\,*} = {\bf p}_{3,\parallel}^{\,*} + {\bf p}_{3,\perp}^{\,*}$, where ${\bf
p}_{3,\parallel}^{\,*}$ and ${\bf p}_{3,\perp}^{\,*}$ are the components parallel
and perpendicular to ${\bf p}^{\,*}_1$, respectively.
Since $V_3$ and $V_4$ are collinear with the beam direction in the effective $W$
approximation valid at high energies,
\begin{eqnarray}
  t \al=\al - \left({\bf p}_1^{\,*}- {\bf p}_{3,\parallel}^{\,*}\right)^2 -
  \left({\bf p}_{3,\perp}^{\,*}\right)^2 \nonumber\\
  \al=\al -{\bf p}_\text{cm}^2 (1-\cos\theta)^2 - {\bf p}_T^2,
  \label{eq:t2}
\end{eqnarray}
where ${\bf p}_T$ is the transverse momentum of particle $V_3$ in the laboratory
frame.
To obtain the second equality, we have used the fact that the perpendicular
component of ${\bf p}_3$ is invariant under the Lorentz boost from the laboratory
frame to the cm frame. From Eqs.~\eqref{eq:t1} and \eqref{eq:t2}, we
obtain the following relations for ${\bf p}_T^2$ assuming $m_A=m_C$ and $m_B=m_D$,
\begin{eqnarray} \label{Ptandt}
  {\bf p}_T^2 = {\bf p}_\text{cm}^2 \sin^2\theta = -t \left( 1 + \frac{t}{4{\bf p}_\text{cm}^2 }
  \right) .
\end{eqnarray}
For the case that all the particles are massless, then ${\bf p}_\text{cm}=\sqrt{s}/2$,
and
\begin{equation}
  {\bf p}_T^2 = \frac{s}{4}\sin^2\theta = \frac{t\,u}{s},
\end{equation}
where we have used $s+t+u=0$.
\end{appendix}

\medskip

\end{document}